\documentclass[twocolumn,aps,pra,superscriptaddress,a4paper,balancelastpage]{revtex4-2}
\usepackage[normalem]{ulem}
\usepackage{amsmath}
\usepackage{amssymb}
\usepackage{gensymb}
\usepackage{bm}
\usepackage{graphicx}
\usepackage{float}
\usepackage{subfigure}
\usepackage[dvipsnames]{xcolor}
\usepackage{placeins}
\usepackage{braket}
\usepackage{bbold}
\usepackage{ulem}
\usepackage[colorlinks, linkcolor=blue, citecolor=blue, urlcolor=blue, breaklinks=red]{hyperref}

\usepackage{bbm}

\usepackage{lipsum}
\usepackage{physics}
\usepackage{dsfont}

\definecolor{Mycolor1}{HTML}{44aa99}
\definecolor{Mycolor2}{HTML}{cc6677}
\usepackage{orcidlink}

\begin{document}

\title{Basis-independent coherence and its distribution in de Sitter spacetime}

\author{Samira Elghaayda~\!\!\orcidlink{0000-0002-6655-0465}}
\email{samira.elghaayda-etu@etu.univh2c.ma}
\affiliation{LMHEP, FSAC, Hassan II University,  Casablanca, Morocco.}

\author{Atta ur Rahman~\!\!\orcidlink{0000-0001-7058-5671}}
\email{attazaib5711@gmail.com}
\affiliation{SPC, University of Chinese Academy of Sciences,  Beijing, China}

\author{Mostafa Mansour~\!\!\orcidlink{0000-0003-0821-0582}}
\email{mostafa.mansour.fsac@gmail.com}
\affiliation{LMHEP, FSAC, Hassan II University, Casablanca, Morocco.}

\begin{abstract}
Quantum coherence in curved spacetime offers a fresh window into the interplay between gravity, thermality, and quantum resources. While previous work has shown that Markovian evolution can generate entanglement and other nonclassical correlations in de Sitter backgrounds, the basis-dependent nature of coherence has so far limited its unambiguous interpretation. Here, we introduce a basis-independent framework to quantify not only the total coherence of two comoving detectors, but also its collective and localized contributions, and we trace how each of these decomposed measures varies with the inverse of Gibbons–Hawking temperature. By treating the detectors as open quantum systems interacting with a massless scalar field in the Bunch–Davies and squeezed $\alpha$-vacua, we find that non-thermal squeezing substantially enhances extractable coherence, even under strong thermal effects. Our results demonstrate how basis-independent coherence in de Sitter spacetime can serve as a robust resource for relativistic quantum information protocols. 
\end{abstract}


\maketitle

\section{Introduction}
De Sitter space is notable for its maximal symmetry and its role in the inflationary universe. The eternal Hubble horizon triggers primordial curvature fluctuations, contributing to the cosmic microwave background (CMB) signature observed today \cite{gibbons1983very,lin2024entanglement,wu2023would,wu2022gaussian}. In the static patch of de Sitter space, a cosmic event horizon behaves like black holes, exhibiting thermal properties similar to Hawking radiation.  The complete isometry of De Sitter space supports various vacua that adhere to CPT invariance, each identified by a superselection parameter $\alpha$ \cite{mottola1985particle,allen1985vacuum}, exhibiting nontrivial ultraviolet and infrared properties, including acausal correlations \cite{mottola1985particle,allen1985vacuum,higuchi2018vacuum,bousso2002conformal}. Among these vacua, the Bunch-Davies vacuum transitions to the Minkowski vacuum as $\alpha$ approaches $-\infty$ and the cosmological constant approaches zero. Other vacua can be interpreted as squeezed states derived from the Bunch-Davies vacuum. There is ongoing debate regarding the consistency of interacting field theory in any specific $\alpha$-vacuum \cite{einhorn2003interacting,goldstein2003note,collins2003fate,de2005alpha} and its implications for understanding new physics at the Planck scale in the early universe \cite{danielsson2002inflation,danielsson2002note,agocs2020quantum} or in holographic contexts \cite{bousso2002conformal,danielsson2002consistency,danielsson2019quantum}.

A powerful method for probing the behavior of quantum fields in curved spacetime is the Unruh-DeWitt detector (UDW) \cite{unruh1976notes,dewitt1979quantum,li2025does}, a localized two-level quantum system that interacts with quantum fluctuations in its environment. In the context of detector-field dynamics, the detector’s response function quantifies the transition rate between energy levels per unit of proper time, as dictated by the field’s propagators in the given spacetime background \cite{birrell1984quantum}. For instance, in flat spacetime, an accelerating UDW which is analyzed within the Rindler coordinate framework, registers a thermal response known as Unruh radiation \cite{crispino2008unruh}. Similarly, near a stationary black hole, a detector immersed in the Hartle-Hawking \cite{candelas1980vacuum} vacuum experiences excitations corresponding to a thermal spectrum with a Planckian distribution, encapsulating the Hawking effect \cite{hawking1975particle}. In a de Sitter space, it has been demonstrated that a comoving UDW detects thermal radiation at a temperature $T_H = H/2\pi$, with $H$ is the Hubble constant \cite{gibbons1977cosmological}. This Gibbons-Hawking effect persists across a wide range of field-curvature couplings and applies to different types of quantum fields, underscoring its broad universality \cite{garbrecht2004unruh}.  

Since the inception of quantum coherence theory, the basis-dependent nature of coherence has been a topic of ongoing debate \cite{hu2017maximum,yin2022basis,radhakrishnan2019basis,radhakrishnan2016distribution,du2024basis,jin2024basis}. Unlike other quantifiers of quantum states, such as entanglement or quantum discord, which are typically basis-independent and considered intrinsic properties, coherence depends on the choice of basis, introducing subjectivity into its characterization. To address this issue, various approaches have been proposed, including optimization over all possible bases \cite{yu2017quantum,luo2017quantum} and using the maximally mixed state as the reference incoherent state \cite{yao2016frobenius,horodecki2003reversible,singh2015maximally,yu2016total,streltsov2018maximal}. However, there is currently no consensus on the most suitable framework for quantifying coherence or when to use basis-dependent versus basis-independent measures. Clarifying the distinctions between these methodologies is essential for the effective and consistent application of quantum coherence theory.

Beyond the entanglement associated with quantum fields, it is fascinating to explore the possibility of harnessing a broader sense of quantumness from the quantum vacuum. It was suggested that quantum coherence can be extracted even with a single  detector \cite{kollas2020field}. Under specific initial energy levels of the background field and particular interaction durations, a detector moving with uniform acceleration can extract more coherence from the field state than a stationary detector. In other words, under certain conditions, the rate of coherence loss may be slower for a moving detector compared to a stationary one.

The goal of this paper is to investigate how basis-independent quantum coherence and its spatial distribution behave in de Sitter spacetime, serving as fundamental indicators of quantumness in a relativistic context. By analyzing these measures for Unruh–DeWitt detectors, we aim to uncover how spacetime curvature, horizon effects, and thermalization influence the persistence and localization of quantum coherence. This approach provides a consistent and observer-independent framework for probing quantum features in curved spacetime and contributes to a deeper understanding of quantum information in gravitational settings.

The manuscript is structured as follows. In section \ref{sec1} we review the UDW detector framework in de Sitter spacetime, derive the Kossakowski matrix, and obtain the asymptotic thermal density matrix. Section \ref{sec4} introduces the basis-independent coherence measures and details their computation for two detectors in both Bunch–Davies and $\alpha$-vacua. In Section \ref{sec5}, we present our numerical analysis of spatial coherence profiles, vacuum-squeezing effects, and temperature dependence, highlighting the roles of curvature and horizon structure. Finally, Section 5 summarizes our main findings.

\section{UDW detectors in de Sitter space \label{sec1}}
This section presents the basics of UDWs, the features of their reduced dynamics, and the $\alpha$-vacua in de Sitter space, along with the notation for our analysis
\subsection{Master equation of UDW detectors }
We examine two detectors that interact locally with a massless scalar field $\varPhi(x)$ along their worldlines. The classical trajectory of each detector, denoted as $x(t)$, is parameterized by its proper time $t$. The total Hamiltonian of the system, incorporating the detectors and the quantum scalar field, is expressed as 
\begin{equation}
H=H_{sys}+H_\varPhi+H_{int},
\end{equation}
$H_{\rm sys}$ is the Hamiltonian of two non-interacting detectors, with each two-level detector described by a $2 \times 2$ matrix \cite{elghaayda2023entropy,yu2011open,feng2015uncertainty,elghaayda2024distribution}.

\begin{equation}
	H_{sys}=\frac{\omega}{2}\left( {\hat{s}_{3}^{A}}\otimes\hat{s}_{0}^{B}+\hat{s}_{0}^{A}\otimes{\hat{s}_{3}^{B}}\right)\equiv \frac{\omega}{2}\Gamma_3,
\end{equation}
The symmetrized bipartite operators are introduced as  
\begin{equation}
\Gamma_i={\hat{s}_{i}^{A}}\otimes\hat{s}_{0}^{B}+\hat{s}_{0}^{A}\otimes{\hat{s}_{i}^{m}},
\end{equation}
where ${\hat{s}_{i}^{k}}$ $(i=1,2,3)$ denote the Pauli matrices, the superscript $m=\{A,B\}$ distinguishes the two detectors, and $\omega$ represents their energy spacing. 
The Hamiltonian $H_\varPhi$ corresponds to the free massless scalar fields $\varPhi(x)$, which satisfy the KG equation $\Box \varPhi(x)=0$ in de Sitter spacetime. 
The covariant d’Alembertian, $\Box = g^{\mu\nu}\nabla_\mu\nabla_\nu$, depends on the selected coordinate system~\cite{yu2011open}.  

The interaction Hamiltonian $H_{int}$ describes the coupling between the system and the scalar field, modeled as an electric dipole interaction:
\begin{equation}
H_{int}=\lambda\left[(\hat{s}_2^{A}\otimes\hat{s}_{0}^{B})\varPhi(t,\mathbf{x}^{A})+(\hat{s}_{0}^{A}\otimes\hat{s}_2^{B})\varPhi(t,\mathbf{x}^{B})\right],
\end{equation}
where $\lambda$ is a small, dimensionless coupling constant.  
To investigate the dynamical evolution of the reduced density matrix of the system, we define  $ \eta(t)=\mathrm{Tr}_\varPhi[\eta_{tot}(t)],$ with $t$ the proper time along their worldlines. 
We assume an initially separable state of the form 
\begin{equation}
\eta_{tot}(0)=\eta(0)\otimes|0\rangle\langle 0|,
\end{equation}
Here, $|0\rangle$ represents the vacuum configuration of the field $\varPhi(x)$, while $\eta(0)$ specifies the initial density matrix of the detectors.
Because the joint setup of the UDW detectors and the field forms an isolated quantum system, its global density operator evolves under the von Neumann equation,
\begin{equation}
i\dot{\eta}_{\text{tot}}(t) = [H, \eta_{\text{tot}}(t)].
\end{equation}
In the weak-coupling approximation, one can nonetheless derive the effective open-system dynamics by performing a partial trace over the scalar field. 
This leads to a Markovian evolution governed by the Kossakowski–Lindblad master equation\cite{gorini1976completely,lindblad1976generators}
\begin{equation}\label{mase}
	\frac{\partial \eta(t)}{\partial t}=\frac{1 }{i}\left[ H_{eff} , \eta(t)\right] +\mathcal{L}\left[\eta(t)\right],
\end{equation}
with 
\begin{equation}
H_{eff}=H_{sys}- \frac{i }{2} \sum_{m, n = 1}^2 \Omega^{(mn)}_{ij} \hat{s}^{(m)}_i \hat{s}^{(n)}_j,   
\end{equation}
The non-unitary contribution to the detectors’ dynamics is described by  
\begin{equation}
	\mathcal{L}\left[\eta\right]=\sum_{\substack{i,j=1,2,3 \\ k,l=a,b}} 
	\frac{\Omega_{ij}^{(mn)}}{2}\left[ 2\hat{s}_j^{(k)}\eta\hat{s}_i^{(l)}-\left\lbrace \hat{s}_i^{(k)}\hat{s}_j^{(l)},\eta \right\rbrace \right],
\end{equation}
where $\Omega_{ij}^{(mn)}$ are the Kossakowski matrices, determined through the Fourier transform of the scalar-field Wightman functions
\begin{equation}
Y^{(mn)}(\Delta t)=\langle 0| \varPhi(t,x^{(m)}) \varPhi(t',x^{(n)})|0 \rangle,
\end{equation}
with Fourier transform
\begin{equation}
	\mathcal{Z}^{(mn)}(\omega)=\int_{-\infty}^{+\infty} Y^{(mn)}(\Delta t)  ~~e^{i\omega\Delta t}\, d\Delta t\,.
\end{equation}
The indices $m,n={A,B}$ label the detectors and $\Delta t=t-t'$. For two detectors, $Y^{(BB)}=Y^{(AA)}$ and $Y^{(BA)}=Y^{(AB)}$, leading to $\mathcal{Z}^{(BB)}=\mathcal{Z}^{(AA)}\equiv \mathcal{Z}_{0}$ and $\mathcal{Z}^{(BA)}=\mathcal{Z}^{(AB)}$.

The equation (\ref{mase}) thus governs the asymptotic equilibrium state of the detectors at late times, resulting from the interplay between environmental dissipation in de Sitter space and quantum correlations generated by the Markovian dynamics \cite{benatti2004entanglement, benatti2003environment}.  
In the two-detector system, the initial separation $L \equiv |\mathbf{x}^A - \mathbf{x}^B|$ controls correlation generation, as the Kossakowski matrices explicitly depend on this distance. 
Typically
\begin{equation}
\mathcal{Z}^{(BA)}=\mathcal{Z}^{(AB)}\equiv\mathcal{Z}(\omega,L)= f(\omega,L) \mathcal{Z}_{0}(\omega),
\end{equation}
Here, $f(\omega,L)$ is even in $\omega$ \cite{yu2011open,hu2013quantum}. Correlations grow as $L$ decreases and disappear at infinite separation \cite{benatti2005controlling}. 
Moreover, it was shown in \cite{benatti2010entangling} that below a critical distance $L$, correlations can survive asymptotically in the final equilibrium states despite environmental dissipation.  
For this reason, one may restrict to sufficiently small $L$ and focus on the effect of decoherence on equilibrium properties.  
In this regime, all Kossakowski matrices coincide \cite{hu2011entanglement}:  
\begin{equation}
\Omega_{ij}^{(AA)}=\Omega_{ij}^{(BB)}=\Omega_{ij}^{(AB)}=\Omega_{ij}^{(BA)},
\end{equation}
and take the form
\begin{equation}
	\Omega_{ij}=\kappa_+\delta_{ij}-i\kappa_- \epsilon_{ijk}\delta_{3k}+\tau\delta_{3i}\delta_{3j},
\end{equation}
with
\begin{equation}\label{gam}
	\kappa_{\pm}=\tfrac{1}{2}\left( \mathcal{Z}_{0}(\omega)\pm \mathcal{Z}(-\omega)\right), 
	\quad \tau=\mathcal{Z}_{0}(0)-\kappa_+.
\end{equation}
Through the treatment of the master equation (\ref{mase}), one obtains the equilibrium reduced density matrix of the two detectors, expressible in the Bloch representation \cite{benatti2004entanglement, benatti2003environment}.
\begin{equation}\label{e10}
\eta(t)=\frac{1}{4}\left[\hat{s}_{0}^{A}\otimes\hat{s}_{0}^{B}
+\sum_{j=1}^{3}\rho_{j}\Gamma_{j}
+\sum_{i,j=1}^{3}\rho_{ij}\hat{s}_{i}^{A}\otimes\hat{s}_{j}^{B}\right],
\end{equation}
where
\begin{equation}\label{eq11}
	\rho_{j}=-\frac{R}{3+R^2}(\tau+3)\delta_{3j},
\end{equation}
and
\begin{equation}\label{eq12}
\rho_{ij}=\frac{1}{3+R^2}\Big[R^2(\tau+3)\delta_{3i}\delta_{3j}+(\tau-R^2)\delta_{ij}\Big].
\end{equation}
Here, $R = \kappa_-/\kappa_+$ is set by the system evolution, while the parameter 
\[
\tau=\sum_{ii}\eta_{ii}(0)
\] 
is a conserved quantity depending on the initial state.  
It satisfies $\tau\in[-3,1]$ to ensure the positivity of $\eta(0)$.  

\subsection{Dynamics of UDW within $\alpha$-vacua}
We consider freely falling UDW detectors in de Sitter spacetime, weakly interacting with a massless scalar field. In global coordinates $(r, X, \theta, \phi)$, the metric is
\begin{equation}\label{cord}
ds^2 = dr^2 - \frac{\cosh^2(Hr)}{H^{2}}\Big[dX^2 + \sin^{2}X(d\theta^{2} + \sin^{2}\theta, d\phi^{2})\Big],
\end{equation}
with $H$ setting both $\Lambda = 3H^2$ and curvature radius $l=H^{-1}$.

For $L \ll l$, equilibrium correlations no longer depend on detector separation, leading to the steady states of Eqs.~(\ref{e10})–(\ref{eq12}). The scalar field vacua considered are the $\alpha$-vacua~\cite{allen1985vacuum}, a CPT-invariant family labeled by $\alpha<0$.\\

Through the Mottola–Allen transformation, one obtains the $\alpha$-vacua $|\alpha\rangle$, a one-parameter class of de Sitter–invariant states. They can be viewed as squeezed excitations built upon the Bunch–Davies (B-D) vacuum.
\begin{equation}
\left|\alpha\right\rangle = \hat{S}(\alpha) \left|BD\right\rangle,
\end{equation}
Here, $\text{Re}(\alpha) < 0$, and $\hat{S}(\alpha)$ is the standard squeezing operator from quantum optics~\cite{danielsson2002consistency,einhorn2003interacting,einhorn2003squeezed,goldstein2003note,collins2004taming}. We consider CPT-invariant $\alpha$-vacua, for which $\alpha$ is real and negative, with $|\alpha|$ representing its magnitude in all figures and analyses. In this case, the scalar field’s Wightman function in an $\alpha$-vacuum can be obtained in terms of the B-D Wightman function $Y^+_{\rm BD}$ as

\begin{equation}\label{sh}
	\begin{array}{cc}
		\begin{aligned}
			Y^+_{\alpha}(x,y)& =\frac{1}{N}\left[ Y^+_{BD}(x,y)+e^{2\alpha} Y^+_{BD}(y,x) \right.\\
			&\left. e^{\alpha}\left(Y^+_{BD}(x,y_A)+ Y^+_{BD}(x_A,y) \right) \right] 
		\end{aligned}
	\end{array},
\end{equation}
In this notation, the subscript $\alpha$ indicates the chosen $\alpha$-vacuum, with $N =- e^{2\alpha} + 1 $ and $x_A$ denoting the antipodal point of $x$. As $\alpha \to -\infty$, the Wightman function $Y^+_{\alpha}(x,y)$ approaches the Bunch–Davies form $Y^+_{\rm BD}(x,y)$, which reproduces the Minkowski two-point function at short distances in the flat-space limit. Additionally, one can employ the relation \cite{bousso2002conformal}
\begin{equation}\label{e13}
	Y^+(x_A,y)= Y^+(x,y_A)=Y^+(t-i\pi),
\end{equation}
By inserting Eq.~(\ref{e13}) into Eq.~(\ref{sh}), one can compute the Fourier transform of $Y^+_{\alpha}(x,y)$ in the non-Bunch–Davies sectors to get
\begin{equation}
	\mathcal{Z}=\frac{ (2\pi)^{-1}\omega(e^{\alpha-\pi \omega}+1)^{2}}{ (1-e^{2\alpha}) (1-e^{-\beta\omega})}.
\end{equation}
The related Kossakowski coefficients are explicitely expressed  in \cite{elghaayda2025quantum}
\begin{equation}\label{she}
	\begin{array}{cc}
		\begin{aligned}
			\Sigma_{+,\alpha}& =\frac{\omega\left[ e^{-\beta\omega}(1+e^{\pi \omega + \alpha})^2 + ~(1+e^{-\pi \omega + \alpha})^{2}\right] }{4\pi ~(1-e^{-\beta\omega})(1-e^{2\alpha})},\\
			\Sigma_{-,\alpha}& =~\frac{\omega\left[ -e^{-\beta\omega}(1+e^{\pi \omega\alpha+})^2 + (1+e^{-\pi \omega + \alpha})^{2}\right] }{4\pi ~(1-e^{-\beta\omega})(1-e^{2\alpha})},\\
			P_{\alpha}& =\frac{ -e^{-\beta\omega}(1+e^{\pi \omega+\alpha})^2 + (1+e^{-\pi \omega + \alpha})^{2} }{e^{-\beta\omega}(1+e^{\pi \omega+ \alpha})^2 + (1+e^{-\pi \omega+\alpha})^{2} }.
		\end{aligned}
	\end{array}
\end{equation}
With fixed detector energy spacing, one finds that $P_{\alpha} \in [-1, +1]$ as $\beta$ varies. Substituting Eq.~(\ref{she}) into Eq.~(\ref{e10}) then yields the final equilibrium state of the UDW detectors, which takes the form
\begin{equation}\label{xs}
	\eta(t)=\left(
	\begin{array}{cccc}
		\eta_{-} & 0 & 0 & 0 \\
		0 & \eta_{22} &\eta_{23}  & 0 \\
		0 & \eta_{23} & \eta_{22} & 0 \\
		0 & 0 & 0 & \eta_{+}
	\end{array}
	\right),
\end{equation}
\begin{equation}
\eta_{\pm}=\frac{(\tau+3)(P_{\alpha}\pm1)^2}{4(P_{\alpha}^2+3)},\quad
	 \eta_{23}=\frac{\tau-P_{\alpha}^2}{2(3+P_{\alpha}^2)},
\end{equation}
\begin{equation}
\eta_{22}=\frac{3-\tau-(1+\tau)P_{\alpha}^2}{4(3+P_{\alpha}^2)}.
\end{equation}
The eigenvalues of the final state in Eq.~(\ref{xs}) are given by
\begin{equation}\label{eid}
\begin{split}
    \mu_1 &= \frac{(1 - P_{\alpha})^2 (\tau + 3)}{4 (P_{\alpha}^2 + 3)}, \quad
    \mu_2 = \frac{(1 + P_{\alpha})^2 (\tau + 3)}{4 (P_{\alpha}^2 + 3)},\\
    \mu_3 &= \frac{(1 - P_{\alpha}^2) (\tau + 3)}{P_{\alpha}^2 + 3}, \quad
    \mu_4 = \frac{1 - \tau}{4}.
\end{split}
\end{equation}

The eigenvectors in the computational basis are given by
\begin{equation}\label{vpxs}
\begin{split}
    |\phi\rangle_1 &= |00\rangle, \quad |\phi\rangle_2 = |11\rangle,\\
    |\phi\rangle_3 &= \frac{1}{\sqrt{2}} (|10\rangle - |01\rangle), \quad
    |\phi\rangle_4 = \frac{1}{\sqrt{2}} (|01\rangle - |10\rangle).
\end{split}
\end{equation}
Prior to further analysis, several notable properties of $\alpha$-vacua should be highlighted.
Firstly, the squeezing inherent to these vacua can markedly lower measurement uncertainties, suggesting that some quantum correlations may remain concealed in their non-Bell-diagonal structure \cite{maldacena2013entanglement,kanno2014entanglement}.
Secondly, $\alpha$-vacua do not generally display thermal characteristics, as Eq.~(\ref{sh}) violates the Kubo–Martin–Schwinger condition unless it coincides with the Bunch–Davies vacuum.
This lack of thermality has motivated investigations \cite{martin2001trans,kaloper2003initial,goldstein2003initial,ashoorioon2014non} treating $\alpha$-vacua as alternative initial states in models of inflation.
Finally, to account for expected modifications to the primordial power spectrum on the order of $\mathcal{O}(H/\Lambda)^2$, $\alpha$ can be interpreted in terms of a fundamental energy scale $\Lambda$, such as the Planck or string scale \cite{danielsson2002inflation,danielsson2002note}.

\section{Measures of quantum coherence}
In this section, we provide an overview about concept of basis-independent measure of quantum coherence using the Jensen–Shannon divergence and its distribution in de Sitter spacetime. This choice is motivated by the need for resource measures that remain physically meaningful across different reference frames and observers. Basis-independent coherence offers a robust and observer-invariant characterization of intrinsic quantumness, making it especially suited for relativistic contexts. By focusing on these measures, we aim to capture the fundamental quantum features of the system in a way that is consistent with the geometric structure of de Sitter spacetime and operationally relevant for quantum information tasks in a gravitational background. 

\subsection{Total, collective, and localized coherence}

Quantum coherence is a fundamental resource in quantum physics and quantum information processing. Radhakrishnan \textit{et al.} proposed a measure based on the quantum Jensen–Shannon divergence \cite{radhakrishnan2019basis,radhakrishnan2016distribution}, defined for a state $\eta$ as
\begin{equation}
\mathrm{C}_{\mathrm{JSD}}(\eta) = \sqrt{ S\Big( \frac{\eta + \eta_d}{2} \Big) - \frac{1}{2} \big[ S(\eta) + S(\eta_d) \big] },
\end{equation}
where $\eta_d$ is the nearest incoherent state obtained by removing off-diagonal elements, making the measure basis-dependent. To remove this dependence, one can use the maximally mixed state $\eta_I = I/d$ as a reference, giving the basis-independent coherence
\begin{align}
\mathrm{C}_\mathrm{T}(\eta) = \sqrt{ S\Big( \frac{\eta + \eta_I}{2} \Big) - \frac{1}{2} \big[ S(\eta) + \log_2 d \big] }.
\end{align}
In multipartite systems, collective coherence accounts for coherence arising from correlations among subsystems:
\begin{align}
C_C(\eta) = \sqrt{ S\Big( \frac{\eta + \pi_\eta}{2} \Big) - \frac{1}{2} \big[ S(\eta) + S(\pi_\eta) \big] },
\end{align}
with $\pi_\eta = \eta_1 \otimes \eta_2 \otimes \dots \otimes \eta_n$ the tensor product of reduced states. Localized coherence isolates the intrinsic coherence of individual subsystems:
\begin{align}
C_L(\eta) = \sqrt{ S\Big( \frac{\pi_\eta + \eta_I}{2} \Big) - \frac{1}{2} \big[ S(\pi_\eta) + \log_2 d \big] },
\end{align}
thereby separating coherence due to subsystem properties from that emerging from inter-subsystem interactions.

\subsection{Global coherence }
We study quantum consonance, a measure of global coherence in multipartite quantum systems. For a general multipartite density matrix, it is defined as \cite{pei2012using}
\begin{align}
C_G(\eta) = \sum_{k_n, l_n} \Big| \eta_{k_n, l_n}^{c} \prod_m (1 - \delta_{k_m, l_m}) \Big|,
\end{align}
where $\eta^{c} = U \eta U^{\dagger}$ is obtained by applying unitary operations that remove the local coherence of $\eta$. Unlike entanglement, quantum consonance can be efficiently computed even for multipartite systems. For a two-qubit state $\eta$, this reduces to
\begin{align}\label{qw}
C_G(\eta) = \sum_{k_1 k_2, l_1 l_2} \Big| \eta_{k_1 k_2, l_1 l_2}^{c} (1 - \delta_{k_1, l_1}) (1 - \delta_{k_2, l_2}) \Big|,
\end{align}
where $\eta_{k_1 k_2, l_1 l_2}^{c} = (U_A \otimes U_B) \eta (U_A \otimes U_B)^{\dagger}$, with $U_A$ and $U_B$ chosen to eliminate the local coherence of the reduced states $\eta_{A/B} = \mathrm{Tr}_{B/A}(\eta)$. The indices $k_1, l_1$ and $k_2, l_2$ label the two-qubit computational basis ${ |00\rangle, |01\rangle, |10\rangle, |11\rangle }$. From Eq.~(\ref{qw}), it follows that quantum consonance only includes the elements corresponding to $|00\rangle\langle 11|$, $|01\rangle\langle 10|$, $|10\rangle\langle 01|$, and $|11\rangle\langle 00|$ of the transformed state $\eta^{c}$. The extension to systems with more qubits, such as four-qubit states, is straightforward.

\section{Results and analysis \label{sec4}}
Our goal in this section is to compute and visualize the behavior of basis-independent quantum coherence and its distribution in de Sitter space, focusing on how it responds to thermal effects and vacuum structure. We begin by analyzing the dependence of coherence on the parameter $\beta$, which is inversely related to the de Gibbons–Hawking temperature, under different vacuum choices. Fig. \ref{figure1} presents total, collective, localized, and global coherence for a pair of closely spaced UDW detectors, plotted as functions of $\beta$ for selected values of the squeezing parameter $|\alpha|$, at fixed detector energy gap $\omega = 1.5$ and proper time $\tau = 0.9$. This setup allows us to isolate how vacuum squeezing modifies the resilience of quantum coherence in the presence of curvature-induced thermal noise.

\begin{widetext}
\begin{minipage}{\linewidth}
\begin{figure}[H]
\centering
\subfigure[]{\label{figure1a}\includegraphics[scale=0.55]{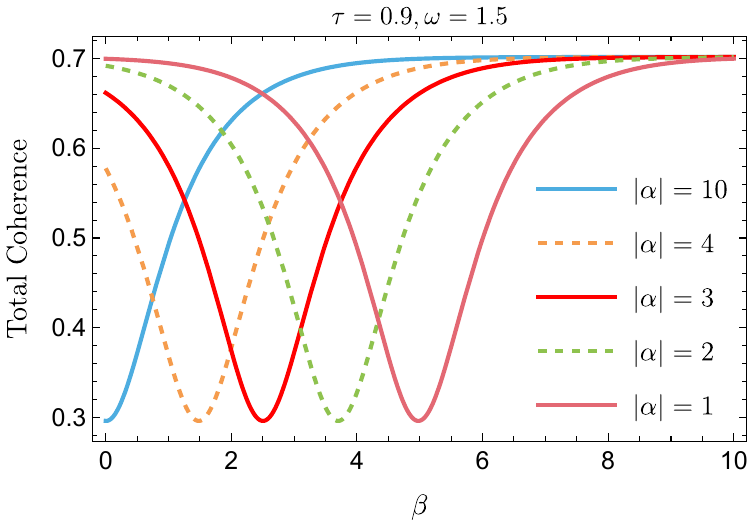}}
\subfigure[]{\label{figure1b}\includegraphics[scale=0.55]{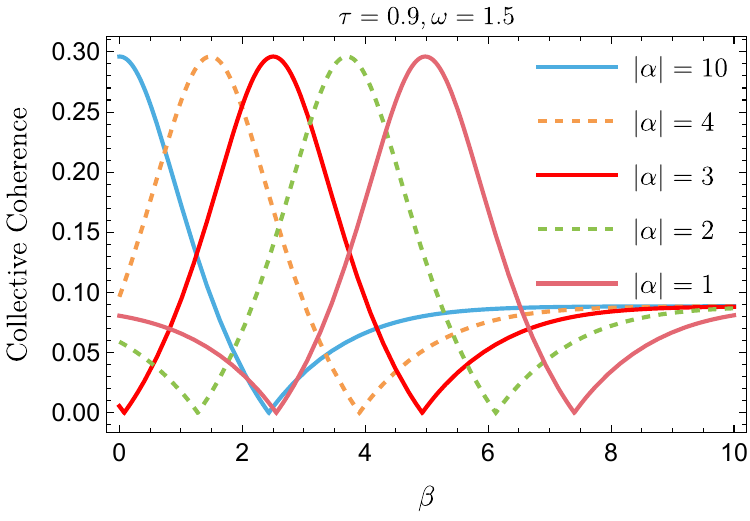}}
\subfigure[]{\label{figure1c}\includegraphics[scale=0.55]{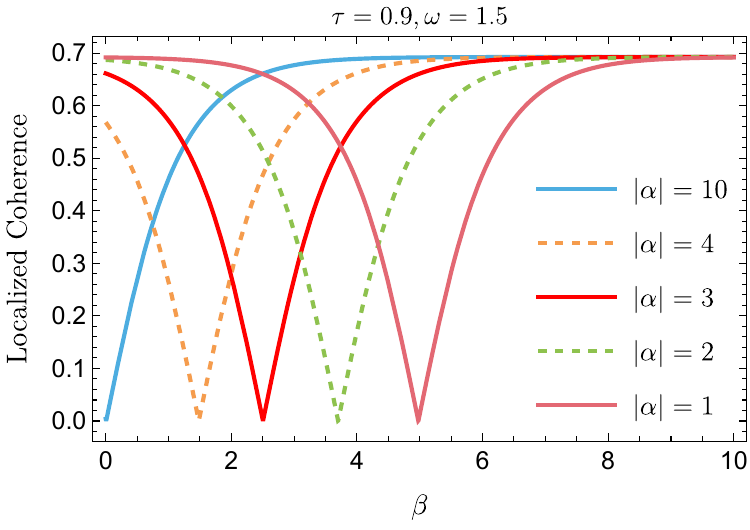}}
\subfigure[]{\label{figure1d}\includegraphics[scale=0.55]{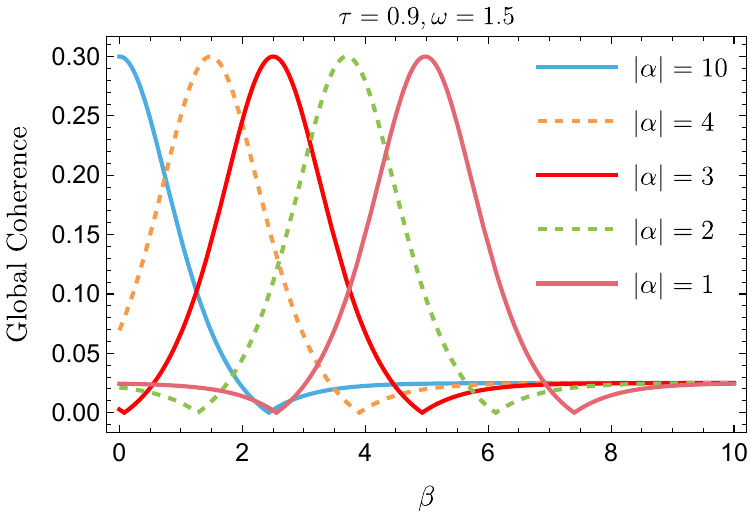}}
\caption{The total $C_T$ \ref{figure1a}, collective $C_C$ \ref{figure1b}, localized $C_L$ \ref{figure1c}, and global coherence $C_G$ \ref{figure1d} associated with the state $\eta(t)$ describing the UDW detectors in de Sitter space, plotted versus $\beta$ for selected magnitudes of $|\alpha|$ with $\omega=1.5$, and $\tau=0.9$.}
\label{figure1}
\end{figure}
\end{minipage}
\end{widetext}
Fig. \ref{figure1} demonstrates the dynamics of the coherence quantifiers strongly governed by the parameter $|\alpha|$ against a specific range of $\beta$. For small to moderate value ($|\alpha| = \{1,2,3,4\}$), the  $C_T$ \ref{figure1a} and $C_L$ \ref{figure1c} exhibit a characteristic U-shaped profile: they initiate from a relatively high value, decrease to a well-defined minimum at $\beta_{\rm min}(|\alpha|)$, and subsequently increase to stabilize at an asymptotic plateau for large $\beta$. In contrast, the $C_C$ \ref{figure1b} and  $C_G$ \ref{figure1d} display a more complex non-monotonic behavior. They show an initial dip at small $\beta$, followed by a rise to a pronounced maximum at $\beta_{\rm max}(|\alpha|)$, an optimality effect, before decaying and then slightly recovering toward their plateau values. The case $|\alpha| = 10$  exhibits markedly distinct behavior, signaling a transition in the physical regime. Here, $C_T$ and $C_L$ increase almost monotonically from a low initial value to the plateau, while $C_C$ displays only a shallow dip at small $\beta$ followed by a gradual rise, entirely lacking the pronounced intermediate peak observed for smaller $|\alpha|$. This distinction stems fundamentally from the changing structure of the quantum vacuum. For large $|\alpha|$ (e.g., $|\alpha| = 10$), the $\alpha$-vacuum approaches the thermal Bunch--Davies limit, where non-thermal correlations induced by vacuum squeezing are strongly suppressed. Consequently, the system dynamics are dominated by thermal decoherence effects from the Gibbons-Hawking radiation, leading to the initial suppression of coherence and its subsequent recovery at lower temperatures (larger $\beta$). In contrast, for smaller $|\alpha|$, the significant squeezing inherent in the vacuum injects substantial non-thermal correlations into the detector system. This enhances the initial coherence but also renders it more susceptible to dissipation at intermediate temperatures, resulting in the observed extrema. The systematic shift of the minima ($\beta_{\rm min}$ for $C_T$, $C_L$) and maxima ($\beta_{\rm max}$ for $C_C$, $C_G$) to lower $\beta$ as $|\alpha|$ increases provides a clear signature of this competition between squeezing-enhanced coherence and thermal dissipation. In practice, the value $|\alpha| = 10$ lies in a regime sufficiently close to the Bunch--Davies limit to exhibit near-thermal behavior, explaining the qualitative difference of its curves. The positions of the extrema $\beta_{\rm min}$ and $\beta_{\rm max}$, and their dependence on $|\alpha|$, offer a quantitative measure of this cross-over and could be further emphasized in the figure with markers or annotations.

Next, in Figure~\ref{figure2}, we investigate how the detector frequency $\omega$ influences basis-independent coherence in the Bunch–Davies vacuum, fixing $|\alpha| = 2$ and $\tau = 0.9$. Subfigures \ref{figure2a} and  \ref{figure2b} display the total and collective coherence as functions of the Gibbons–Hawking parameter $\beta$ for various energy gaps $\omega$, while \ref{figure2c} and \ref{figure2d} show the corresponding local and distributed coherence components.

\begin{widetext}
\begin{minipage}{\linewidth}
\begin{figure}[H]
\centering
\subfigure[]{\label{figure2a}\includegraphics[scale=0.55]{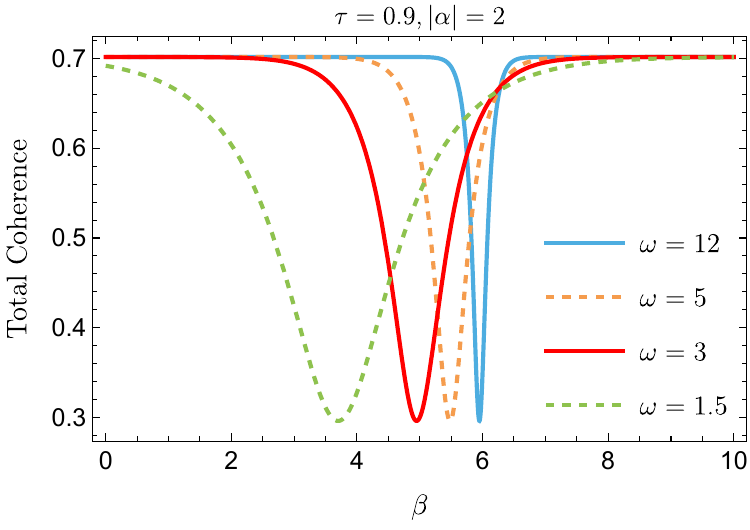}}
\subfigure[]{\label{figure2b}\includegraphics[scale=0.55]{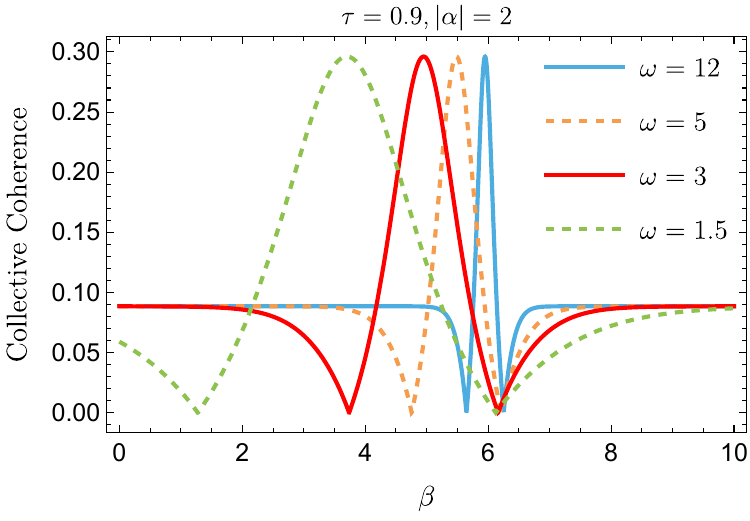}}
\subfigure[]{\label{figure2c}\includegraphics[scale=0.55]{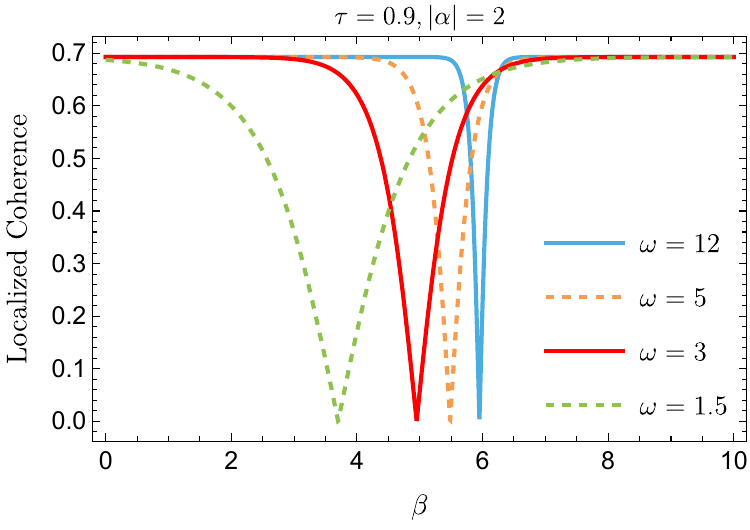}}
\subfigure[]{\label{figure2d}\includegraphics[scale=0.55]{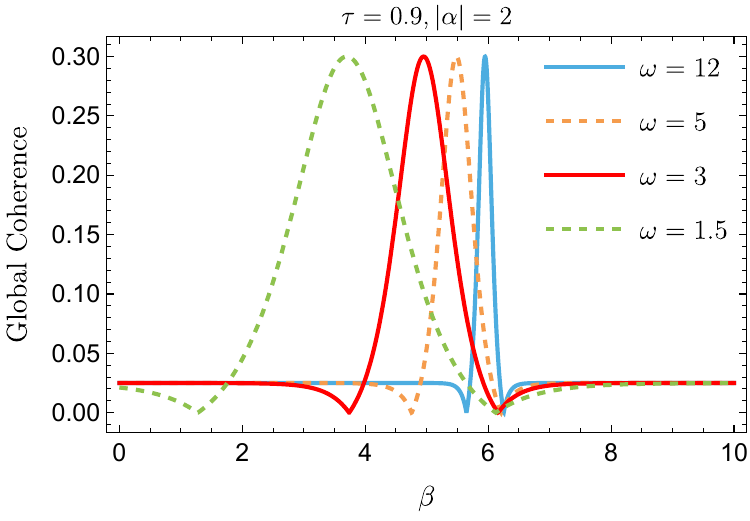}}
\caption{The total $C_T$ \ref{figure2a}, collective $C_C$  \ref{figure2b}, localized $C_L$ \ref{figure2c}, and global coherence $C_G$ \ref{figure2d} associated with the state $\eta(t)$ describing the UDW detectors in de Sitter space, plotted versus $\beta$ for chosen values of $\omega$ with $\left| \alpha\right|=2$, and $\tau=0.9$.}
\label{figure2}
\end{figure}
\end{minipage}
\end{widetext}

Fig. \ref{figure2} displays the asymptotic behavior of the basis-independent coherence components for two UDW in de Sitter spacetime as functions of the inverse temperature $\beta$, with fixed parameters $|\alpha| = 2$, $\tau = 0.9$, and varying detector energy gaps $\omega =\{ 1.5, 3, 5, 12\}$. Fig. \ref{figure2a} shows that the  $C_T$ originates from a higher initial value for all $\omega$, indicating that the initial coherence is predominantly governed by the squeezed nature of the $\alpha$-vacuum and is independent of the detector's internal energy structure. This is followed by a plateau region whose duration scales with $\omega$; larger energy gaps sustain coherence over a broader temperature range. A subsequent sharp decline to a minimum reflects a regime where thermal noise dominates and disrupts coherence. Finally, $C_T$ recovers and converges to a steady-state value, suggesting a balance between vacuum-induced coherence and residual environmental dissipation. Notably, $C_L$ in Fig. \ref{figure2c} closely mirrors the behavior of $C_T$, both qualitatively and quantitatively. This strong correlation indicates that the total coherence is primarily composed of localized, individual detector contributions, especially pronounced in the low-temperature regime.  In contrast, the $C_C$ in Fig. \ref{figure2b} and $C_G$ in Fig. \ref{figure2d} exhibit markedly different behavior. These non-local coherence measures show significantly smaller magnitudes and display non-monotonic profiles with $\omega$-dependent peaks at intermediate temperatures. This distinct behavior underscores that non-local coherences are more sensitive to the detailed structure of field correlations in the $\alpha$-vacuum. Their enhancement within specific temperature windows indicates that collective quantum effects can be optimized by tuning the detector energy gap relative to the environmental temperature. Therefore, the results collectively demonstrate a decoupling between local and non-local coherence mechanisms. While the total coherence is dominated by localized contributions and exhibits universal initial behavior, the collective and global coherences are governed by distinct physical processes and can be selectively enhanced in specific parameter regimes, highlighting the rich interplay between vacuum squeezing, thermal noise, and detector properties in de Sitter space.

\begin{widetext}
\begin{minipage}{\linewidth}
\begin{figure}[H]
\centering
\subfigure[]{\label{figure3a}\includegraphics[scale=0.45]{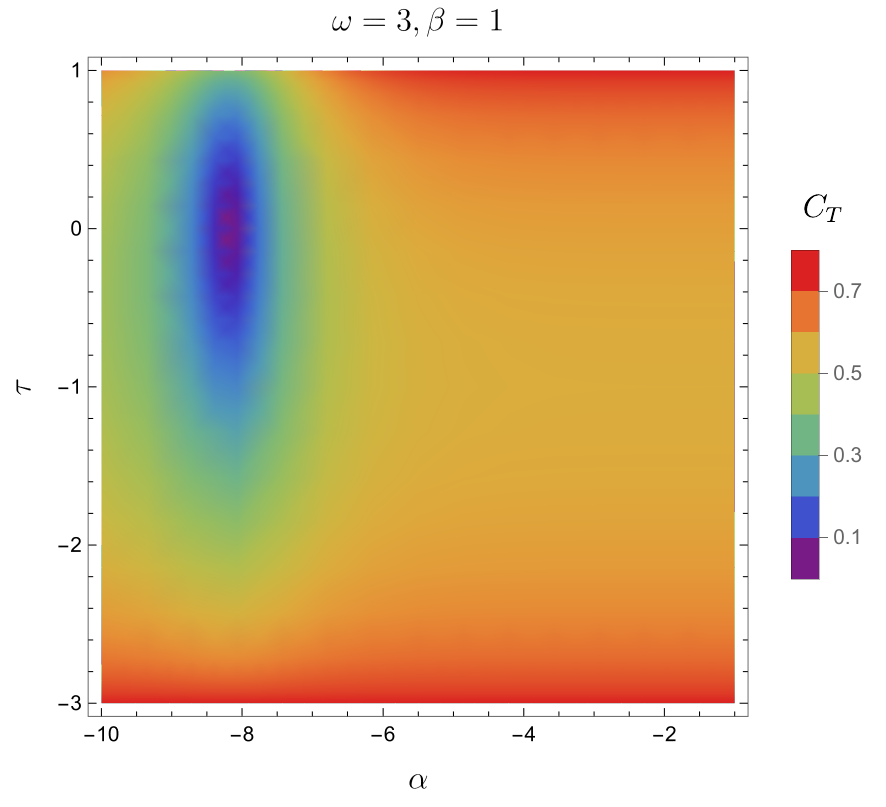}}
\subfigure[]{\label{figure3b}\includegraphics[scale=0.45]{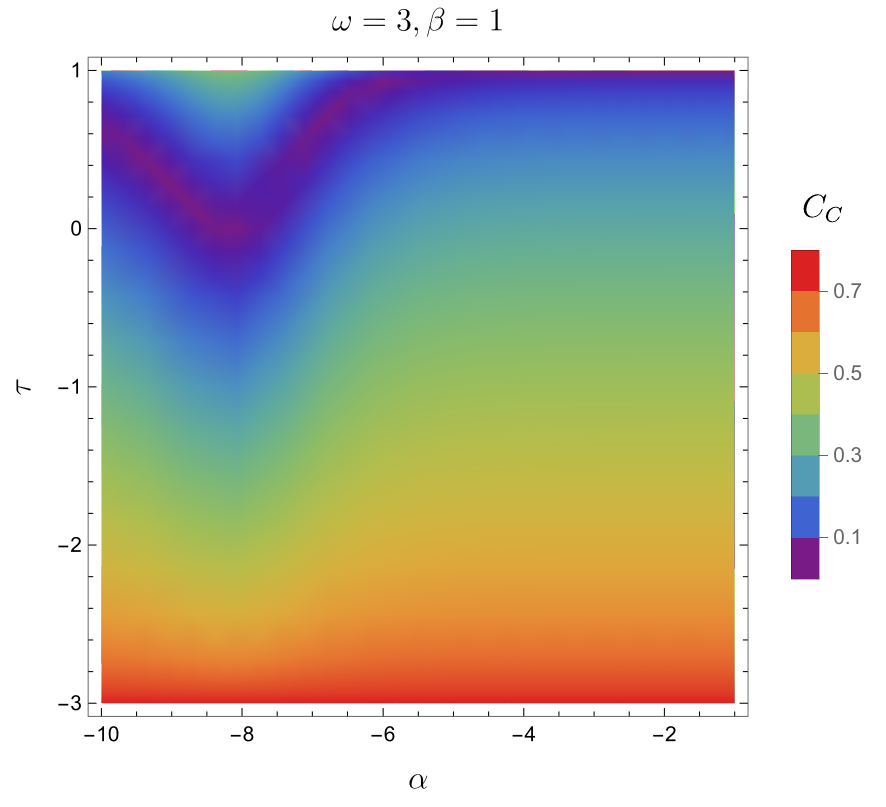}}
\subfigure[]{\label{figure3c}\includegraphics[scale=0.45]{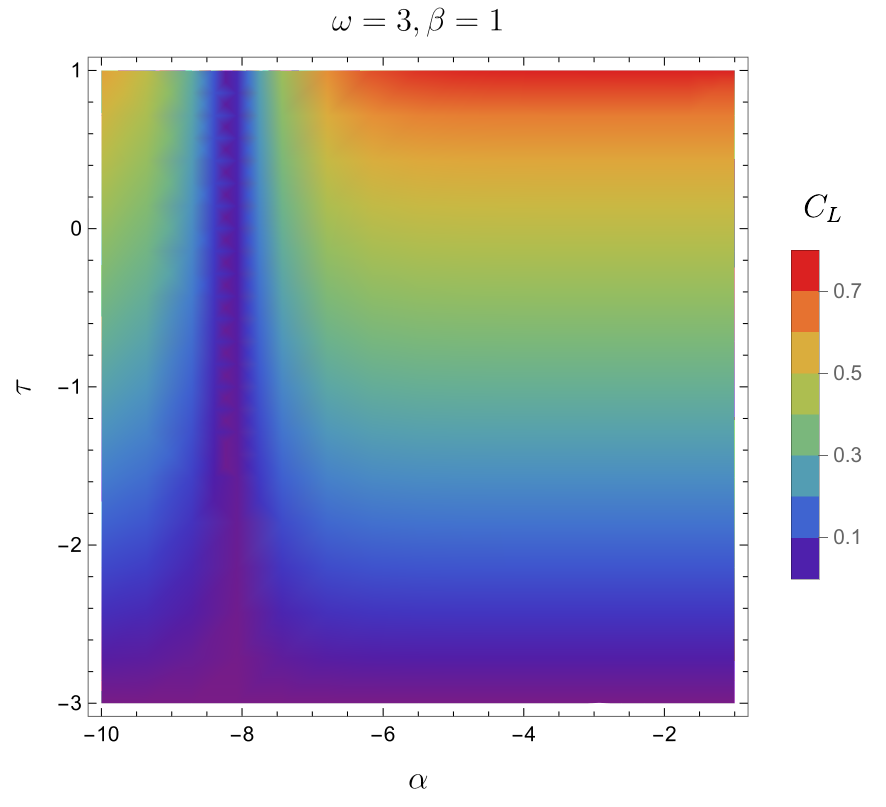}}
\subfigure[]{\label{figure3d}\includegraphics[scale=0.45]{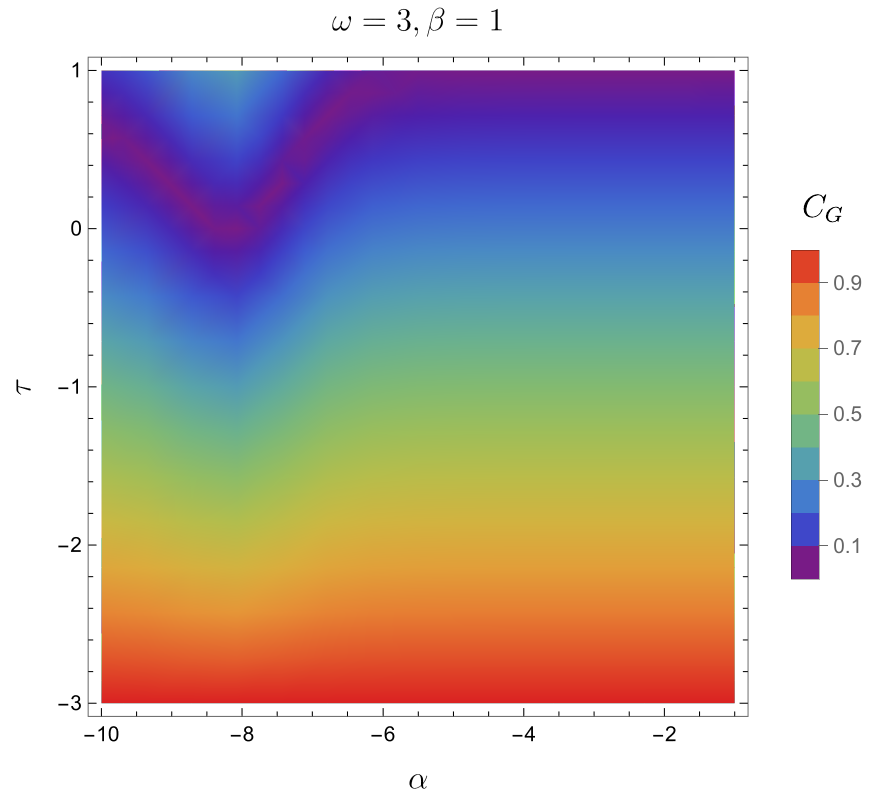}}
\caption{The total $C_T$ \ref{figure3a}, collective $C_C$ \ref{figure3b}, localized $C_L$ \ref{figure3c}, and global coherence $C_G$ \ref{figure3d} associated with the state $\eta(t)$ describing the UDW detectors in de Sitter space, as a function of $\tau$ and $\alpha$ with $\omega=3$, and $\beta=1$.}
\label{figure3}
\end{figure}
\end{minipage}
\end{widetext}
The evolution of quantum coherence measures for the asymptotic state $\eta$, detailed in Fig. \ref{figure3}, provides insight into the distribution of basis-independent coherence in de Sitter space and their dependence on the vacuum structure and initial conditions. The behavior of the $C_T$, $C_C$, $C_L$, and $C_G$ reveals a distinct hierarchy governed by the $\alpha$-vacuum and the detector configuration. In this regard, the parameter $\alpha$ acts as a master controller for the environment's quantum properties. As $\alpha$ increases from $-\infty$ (the Bunch-Davies vacuum) towards zero, the squeezed nature of the $\alpha$-vacuum becomes more pronounced. This squeezing, reflected in the parameter $P_{\alpha}$ from Eq.~\eqref{she}, amplifies non-thermal correlations, leading to a significant enhancement of the total coherence $C_T$, followed by $C_C$ and $C_G$. Conversely, as $\alpha \to -\infty$, all coherence measures converge to their thermal equilibrium values, characteristic of a Gibbons-Hawking temperature bath. We find that coherence is not generated from a fully mixed initial state ($\tau = -3$) as it requires an initial bias or purity. For initial states with $\tau > 0$, the $C_C$ and $C_G$ measures vanish, indicating a highly decohered state. In contrast, $C_L$ exhibits behavior opposite to $C_C$ and $C_G$, also becoming negligible for $0 < \tau \leq 1$, similar to $C_T$. Moreover, the clear dominance of $C_C$ over $C_L$ is a direct consequence of the symmetric coupling imposed by our setup (e.g., $L \ll H^{-1}$). This configuration ensures the detectors interact identically with a common field environment, promoting the generation of quantum correlations between them rather than localized excitations. This dynamics is inherent in the structure of the master equation's dissipator $\mathcal{L}[\eta]$, which is built from collective operators. Most significantly, the substantial magnitude and behavior of $C_G$  is the most crucial result. It confirms that the generated correlations are genuinely quantum-mechanical and non-local. The strong correlation between $C_C$ and $C_G$ indicates that a major portion of the collective coherence is encoded in globally coherent off-diagonal elements, which are an essential resource for quantum information protocols. $C_G$ is arguably the most powerful measure for this system because (i) it isolates the genuinely quantum part of the non-local coherence, (ii) it measures a directly usable resource, and (iii) it is highly sensitive to the vacuum parameter $\alpha$, making it an excellent probe for non-thermal effects. Therefore, the $\alpha$-vacuum acts as a tunable non-thermal environment that can enhance quantum resources. The observed profile: high $C_T$ and $C_C$, low $C_L$, and significant $C_G$ is a characteristic signature of correlated qubits in a squeezed bath. This demonstrates the controllability of global quantum coherence in an expanding universe, with $C_G$ serving as the most robust witness of this non-classicality.

\section{Concluding remarks}\label{sec5}
In this work, we have explored the behavior of basis-independent quantum coherence for UDW detectors in de Sitter spacetime, employing total, collective, localized, and global coherence. By analyzing the asymptotic state of two detectors coupled to a massless scalar field in both the Bunch–Davies and $\alpha$-vacua, we have uncovered several features of coherence generation and distribution in the presence of spacetime curvature and thermality.
Our results demonstrate that the $\alpha$-vacua parameter serves as a powerful control knob, tuning the system between thermal and non-thermal regimes. For small $|\alpha|$, the squeezing inherent in the $\alpha$-vacuum injects substantial non-thermal correlations, significantly enhancing the extractable coherence and leading to pronounced non-monotonic behavior in both collective $C_C$ and global $C_G$ coherence measures. In contrast, as $|\alpha| \to \infty$, the system approaches the thermal Bunch–Davies limit, where coherence is initially suppressed by Gibbons–Hawking radiation before recovering at low temperatures.
Notably, we observed a clear decoupling between local and non-local coherence mechanisms. While the totale $C_T$ and localized $C_L$ coherence are dominated by individual detector contributions and exhibit universal initial behavior, the collective and global coherences are highly sensitive to the detailed field correlations and can be selectively enhanced in specific temperature-frequency windows. In particular, the $C_G$, emerges as the most robust and operationally significant measure, as it specifically quantifies the genuinely non-local, multipartite coherence that is directly usable in  relativistic quantum information protocols.
Furthermore, our analysis reveals that the detector energy gap $\omega$ plays a critical role in sustaining coherence over a broad temperature range, with larger gaps extending the coherence plateau and delaying the disruptive Gibbons-Hawking effect. The initial state parameter $\tau$ also influences the amplitude of coherence, underscoring the importance of initial conditions in coherence harvesting.
Finally, this study establishes that quantum coherence can persist and even be enhanced in de Sitter spacetime, despite the decohering effects of curvature and horizon-induced thermality. The basis-independent measures employed here provide a unified and observer-invariant framework for quantifying quantum resources in relativistic settings. These findings pave the way for future explorations of quantum coherence in more general curved spacetimes and its potential applications in relativistic quantum technologies.

\section*{Disclosures}
The authors declare that they have no known competing financial interests.

\section*{Data availability}
No datasets were generated or analyzed during the current study.

\bibliography{references}
\bibliographystyle{unsrt}

\end{document}